\documentclass[10pt,conference]{IEEEtran}

\usepackage{url}
\usepackage{hyperref}
\usepackage[belowskip=0pt,aboveskip=5pt]{subcaption}
\usepackage{listings}
\usepackage{enumitem}
\usepackage{booktabs}
\usepackage{xspace}
\usepackage{makecell}
\usepackage{amsmath}
\usepackage{pifont}
\usepackage{balance}
\usepackage{soul}
\usepackage{microtype}    
\usepackage[linesnumbered,ruled,vlined]{algorithm2e}
\usepackage[noend]{algpseudocode}
\usepackage{amsthm}
\usepackage{comment}
\usepackage{xr}
\usepackage{wrapfig}
\usepackage{color}
\usepackage[most]{tcolorbox}
\usepackage{setspace}
\usepackage{listings}
\usepackage{lipsum}
\usepackage{adjustbox}
\usepackage{cleveref}
\crefformat{section}{\S#2#1#3} 
\crefformat{subsection}{\S#2#1#3}
\crefformat{subsubsection}{\S#2#1#3}
\usepackage{array,multirow,graphicx}
\usepackage{float}
\usepackage{graphicx}
\usepackage{arydshln}
\usepackage{multirow}

\usepackage[linesnumbered,ruled,vlined]{algorithm2e}
\SetKwInput{KwInputs}{Inputs}                
\SetKwInput{KwOutputs}{Outputs}              
\SetKwInput{KwOutput}{Output}              

\def\BibTeX{{\rm B\kern-.05em{\sc i\kern-.025em b}\kern-.08em
    T\kern-.1667em\lower.7ex\hbox{E}\kern-.f125emX}}

\newcommand{\approach}{\textsc{BugFarm}\xspace}
\newcommand{\leam}{LEAM\xspace}
\newcommand{\mubert}{$\mu$BERT\xspace}
\newcommand{\fitrepair}{FitRepair\xspace}
\newcommand{\codebert}{CodeBERT\xspace}
\newcommand{\codet}{CodeT5\xspace}
\newcommand{\natgen}{\textsc{NatGen}\xspace}

\newcommand{\vulgen}{\textsc{VulGen}\xspace}
\newcommand{\bugswarm}{\textsc{BugSwarm}\xspace}
\newcommand{\chatgpt}{ChatGPT\xspace}
\newcommand{\regminer}{RegMiner\xspace}

\newcommand\printpercent[2]{\the\numexpr#1*100/#2\%}
\newcommand{\mybox}[1]{\begin{tcolorbox}[enhanced, frame hidden, boxsep=0pt]\emph{#1}\end{tcolorbox}}
\preto\tabular{\setcounter{rownumbers}{0}}
\newcounter{rownumbers}
\newcommand\rownumber{\stepcounter{rownumbers}\arabic{rownumbers}}

\IEEEoverridecommandlockouts

\begin{document}


\title{Challenging Bug Prediction and Repair Models with Synthetic Bugs}

\author{\IEEEauthorblockN{Ali Reza Ibrahimzada\textsuperscript{1}, Yang Chen\textsuperscript{1}, Ryan Rong\textsuperscript{2}, Reyhaneh Jabbarvand\textsuperscript{1}}
\IEEEauthorblockA{\textit{\textsuperscript{1}University of Illinois Urbana-Champaign}, Urbana, IL, USA \textit{\textsuperscript{2}Stanford University}, Stanford, CA, USA}
\IEEEauthorblockA{\{alirezai,yangc9,reyhaneh\}@illinois.edu \{ryanrong\}@stanford.edu}
}

\maketitle

\begin{abstract}
Bugs are essential in software engineering; many research studies in the past decades have been proposed to detect, localize, and repair bugs in software systems. Effectiveness evaluation of such techniques requires \textit{complex bugs}, i.e., those that are \textit{hard to detect through testing} and \textit{hard to repair through debugging}. From the classic software engineering point of view, a hard-to-repair bug differs from the correct code in multiple locations, making it hard to localize and repair. Hard-to-detect bugs, on the other hand, manifest themselves under specific test inputs and reachability conditions. These two objectives, i.e., generating hard-to-detect and hard-to-repair bugs, are mostly aligned; a bug generation technique can change multiple statements to be covered only under a specific set of inputs. However, these two objectives conflict in the learning-based techniques: A bug should have a similar code representation to the correct code in the training data to challenge a bug prediction model to distinguish them. The hard-to-repair bug definition remains the same but with a caveat: the more a bug differs from the original code (at multiple locations), the more distant their representations are and easier to detect. This demands new techniques to generate bugs to complement existing bug datasets to challenge learning-based bug prediction and repair techniques. 

We propose \approach to transform arbitrary code into multiple hard-to-detect and hard-to-repair bugs. \approach mutates code in multiple locations (hard-to-repair) but leverages attention analysis to only change the least attended locations by the underlying model (hard-to-detect). Our comprehensive evaluation of $435k+$ bugs from over $1.9M$ mutants generated by \approach and two alternative approaches demonstrates our superiority in generating bugs that are hard to detect by learning-based bug prediction approaches (up to 40.53\% higher False Negative Rate and 10.76\%, 5.2\%, 28.93\%, and 20.53\% lower Accuracy, Precision, Recall, and F1 score) and hard to repair by state-of-the-art learning-based program repair technique (28\% repair success rate compared to 36\% and 49\% of \leam and \mubert bugs). \approach is efficient, i.e., it takes nine seconds to mutate a code with no training overhead.

\end{abstract}

\begin{IEEEkeywords}
Bug Generation, Bug Prediction, Interpretation
\end{IEEEkeywords}

\section{Introduction}

Machine learning is the new automation technology, and software engineering is no exception: State-of-the-art software analysis techniques either fine-tune pre-trained models or prompt Large Language Models (LLMs) to automate code-related tasks~\cite{pan2024lost,chen2024neurosymbolic,ibrahimzada2025alphatrans,liu2024codemind}. This demands the existence of high-quality datasets to assess their actual effectiveness. Concerning bug-related tasks, such datasets should include a diverse set of complex bugs, i.e., those that are hard to detect to challenge bug prediction techniques and hard to repair for debugging approaches.   

\begin{figure*}[t]
    \centering
    \includegraphics[width=0.9\linewidth]{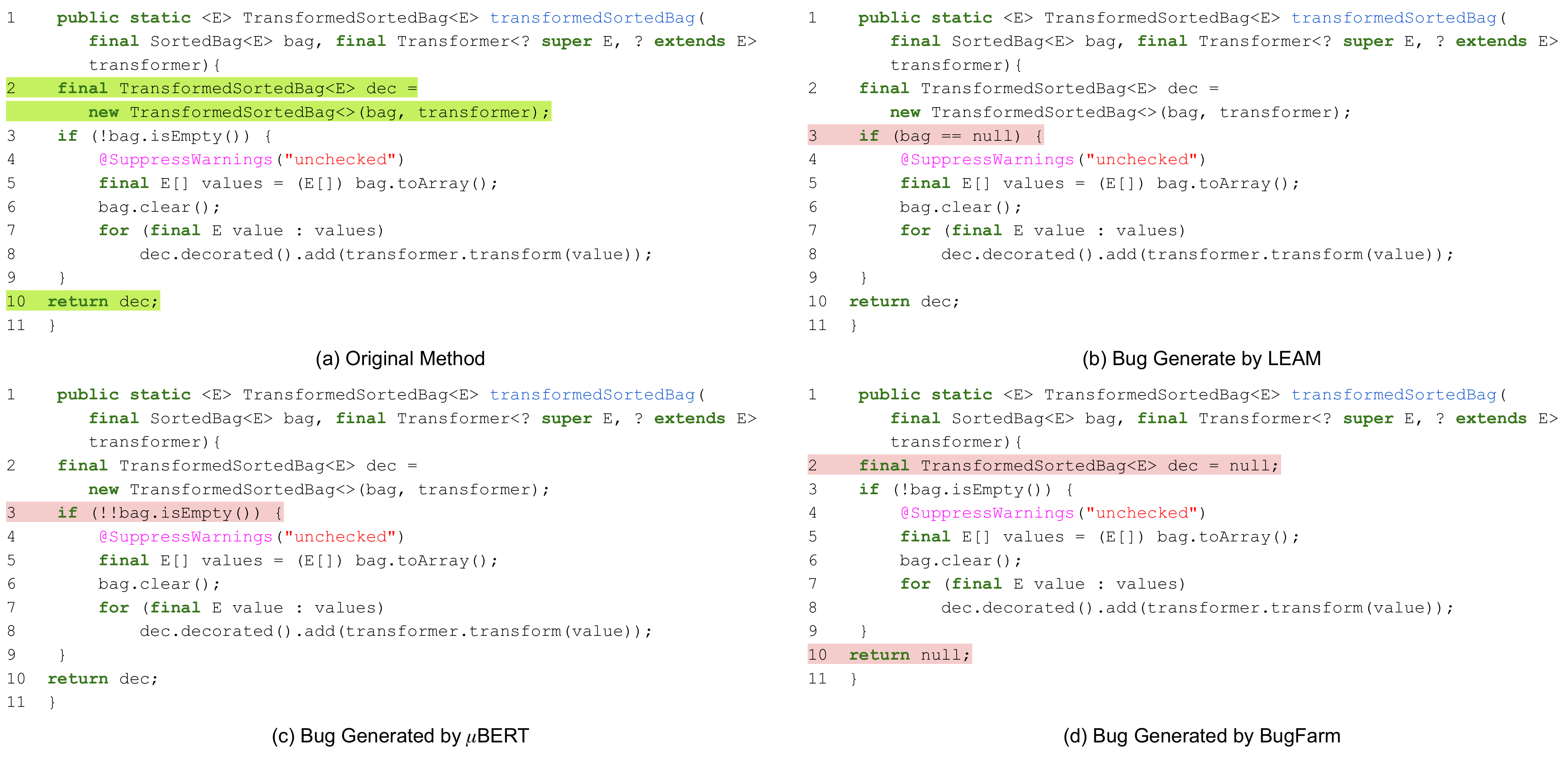}
    \vspace{-5pt}
    \caption{An example showing the bugs generated for the original code (a) using \leam (b), \mubert (c), and \approach (d)}
    \label{fig:example}
    \vspace{-10pt}
\end{figure*}

From the classic software engineering perspective, hard-to-detect bugs often exist at locations reachable only under a particular combination of test inputs or edge cases~\cite{lu2005bugbench,bohme2014corebench}. \emph{However, a similar definition does not stand concerning learning-based bug prediction techniques.} That is because they do not care what inputs trigger the bug in the given code. In contrast, they look for bug patterns in their training/fine-tuning data~\cite{yang2015deep} or to check if the code representation is closer to buggy or correct examples they have seen~\cite{ibrahimzada2022perfect}. As a result, existing real-world bug datasets such as Defects4J~\cite{just2014defects4j}, \bugswarm~\cite{tomassi2019bugswarm}, BugsInPy~\cite{BugsInPy}, \regminer~\cite{song2022regminer}, and ManySStuBs4J~\cite{sstub}, while \emph{necessary} to assess their effectiveness in dealing with real-world bugs, \emph{are not enough} to assess their true effectiveness and rule out data contamination~\cite{samala2020hazards}: To challenge the learning-based bug prediction techniques, one should inject unseen bug patterns such that the code representation of the generated bug and correct code are similar.

Hard-to-repair bugs usually involve multiple statements to challenge debugging techniques, localizing~\cite{li2018enlightened} and fixing automatically. A similar definition is held in learning-based software engineering, but with slight consideration: Modifying a code in multiple locations can change the representation of buggy code, making it easy for bug prediction models to detect it. These two conflicting objectives make the problem of complex bug generation for learning-based techniques challenging. Existing automated mutant generation techniques~\cite{tian2022learning,khanfir2023efficient,jia2009higher}, by default, change a few locations in the code, making their bugs easy to repair and detect.
Some can be configured to modify multiple lines randomly~\cite{khanfir2023efficient} or follow a set of heuristics~\cite{tian2022learning}.
However, this may result in mutants with a different code representation than the correct code, making them easy to detect. 

To advance automated bug generation concerning the evaluation of learning-based bug-related tasks, we propose \approach. For a given code, \approach prompts an LLM to mutate multiple statements. Given that LLMs are potentially creative in generative tasks, leveraging them helps avoid overfitting to a limited number of mutation operators and bug patterns. To ensure that changing many locations does not drastically impact the code representation of mutants, \approach analyzes the attention of the underlying model and instructs LLMs only to change those the model attends least to. The generated code has a similar representation to the original one. Still, it differs within multiple locations, making them hard to detect by bug prediction approaches and hard to repair by debugging techniques. \approach is language-agnostic and can generate bugs for any programming language. Like alternative approaches, the \approach creates unconfirmed bugs and requires a validation process to confirm them as bugs. Our notable contributions are:

\begin{itemize}[leftmargin=10pt]
    \item \textbf{A new perspective into synthetic bug generation:} We propose a novel technique for bug generation concerning the evaluation of the learning-based bug-related tasks. We do not claim our bugs challenge classic bug detection (testing) techniques, as we propose a new definition for hard-to-detect bugs. Our goal is to demonstrate how ignoring properties of learning-based techniques in mutant/bug generation results in subpar assessment of them. 
    The implementation of \approach and all the generated bugs are publicly available~\cite{website}.

    \vspace{3pt}
    \item \sloppy \textbf{Empirical evaluation:} We evaluated \approach and two most recent learning-based mutant generation approaches, \leam~\cite{tian2022learning} and \mubert~\cite{khanfir2023efficient}. Our evaluation of $435k+$ bugs (from over $1.9M$ mutants) generated for $15$ Java projects confirm that compared to others, \approach bugs are hard-to-detect (up to 40.53\% higher False Negative Rate and 10.76\%, 5.2\%, 28.93\%, and 20.53\% lower Accuracy, Precision, Recall, and F1 score) and hard-to-repair (28\% repair success rate compared to 36\% and 49\% of \leam and \mubert bugs) by learning-based techniques. A large-scale human study with $97$ participants and $500$ sampled survived mutants from these techniques shows that \approach mutants are harder to decide buggy or equivalent. Also, generating effective bugs is more efficient using \approach.

    
\end{itemize}

\section{Illustrative Example}
\label{sec:example}
To illustrate the limitations of prior work and present the key ideas behind \approach, we use the code snippets in Figure~\ref{fig:example}. 
There are several ways to inject bugs into original code by adding, deleting, or modifying the $11$ statements in its body (some statements are split into two lines for better presentation). However, the highlighted lines in green show the two least attended statements (details in~\cref{alg:where}).

\begin{figure*}[t]
    \centering
    \includegraphics[width=0.8\textwidth]{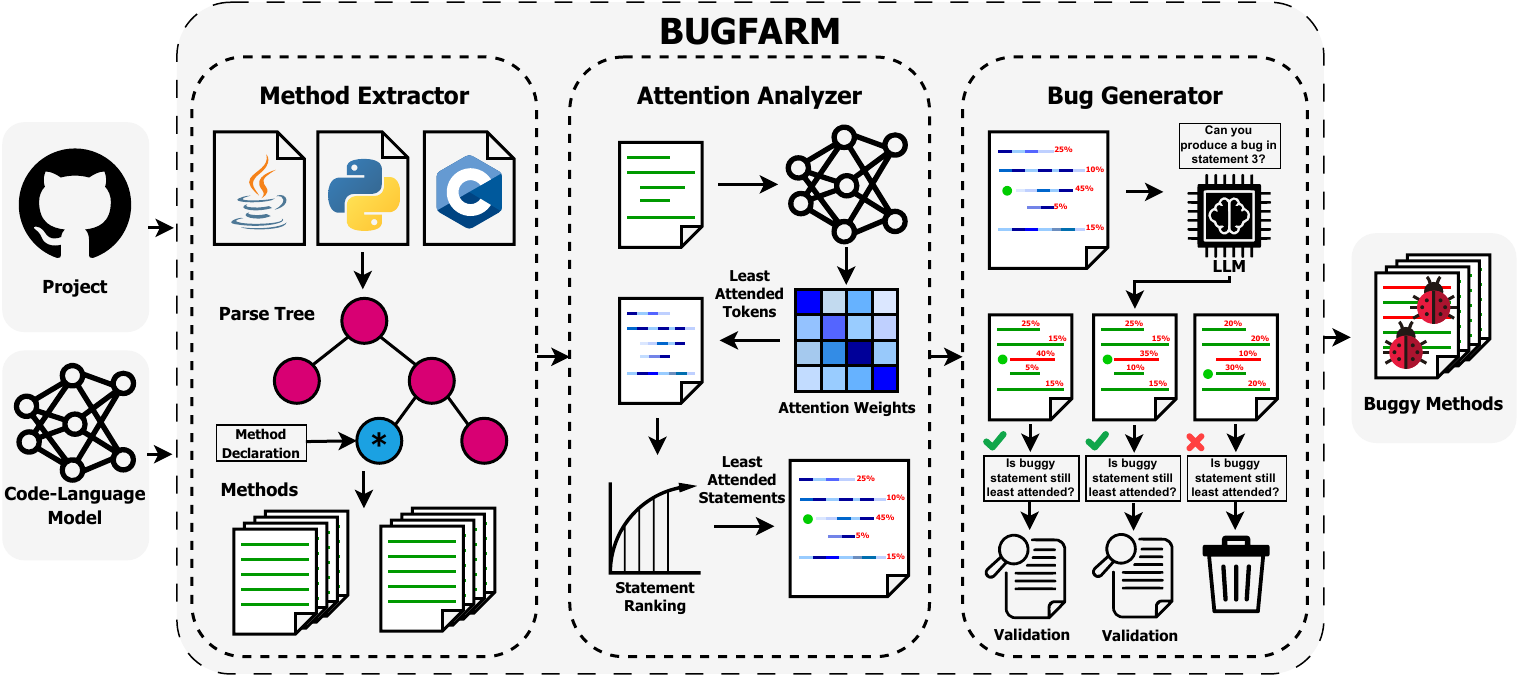}
    \caption{\small{Overview of \approach}}
    \vspace{-10pt}
    \label{fig:overview}
\end{figure*}

Figures~\ref{fig:example}b--\ref{fig:example}d show the bugs generated by three mutant generation tools, \approach, \leam~\cite{tian2022learning}, and \mubert~\cite{khanfir2023efficient} (bugs were confirmed through test execution). 
\leam considers code as a sequence of AST nodes and learns to apply grammar rules to select and modify the code for bug generation. \mubert selects code tokens corresponding to AST nodes, replaces them with a special token \texttt{<mask>}, and asks \codebert to replace them with new tokens for bug generation. \approach identifies the least attended statements and prompts an LLM to perform bug-inducing transformations only on those lines. 
\leam and \mubert generate $19$ and $68$ mutants in total, out of which only $6$ and $1$ are confirmed bugs, respectively. \approach generates the bug in Figure~\ref{fig:example}d. The lines changed to introduce a bug are highlighted in red.

\approach bug is notable from various perspectives. First, it involves multiple statements (hard to localize and repair). Second, both modified statements are among the least attended statements, making it hard for the model to distinguish the bug from the original code (hard to detect). 
\leam and \mubert bugs modify only one line that is not among the least attended statements. As a result, when running against our studied bug prediction model (\cref{evaluation:hard-to-detect}) and program repair model (\cref{evaluation:hard-to-fix}), they were easily detected and repaired. In contrast, the same models failed to detect and repair \approach bug.
\section{Approach Overview}

Figure~\ref{fig:overview} provides an overview of \approach framework consisting of \textit{three} major components: (1) \textit{Method Extractor}, (2) \textit{Attention Analyzer}, and (3) \textit{Bug Generator}. \approach takes a project and a pre-trained code-language model as inputs and extracts the methods through a lightweight static analysis. For each method, it identifies the change location candidates by analyzing the model's attention to individual code tokens. To generate mutants for each method, \approach crafts a prompt to an LLM, including the original code and additional contexts reflecting candidate locations for mutations. 

The \textit{Method Extractor} component takes the input project and builds its corresponding parse tree to extract all the methods in the source files. These methods will be passed as an input to \textit{Attention Analyzer} to identify the candidate locations for the mutation (\cref{subsection:method_extraction}).  

Recent state-of-the-art learning-based software engineering models are all transformer-based, which
rely on attention for neural code representation. So, to identify the candidate locations to inject unnoticeable bugs, \textit{Attention Analyzer} extracts the corresponding attention of the model to code tokens and identifies those with the lower contribution in the code representation, i.e., those with less attention weight values. Bug locations in \approach are at the statement level; thereby, \approach ranks the input method's statements based on the $\#T_{LA}/\#T$ values and chooses the ones with the lowest value. Here, $T_{LA}$ is the number of least attended tokens, and $T$ denotes the total number of tokens in the statement (\cref{subsection:where}).

Finally, \textit{Bug Generator} component takes the list of location candidates and creates a prompt consisting of natural language instruction, location candidates, and the original code. 
For the generated mutants, \approach computes the extent to which the changes impact the model's attention,
and selects mutants with negligible impact on the code representation. It also discards the syntactically incorrect bugs for the sake of quality (\cref{subsection:how}).

\section{\approach}

In this section, 
we first discuss how \approach parses the subjects and extracts methods as inputs for the \textit{Attention Analyzer} component. Next, we will provide a background on the attention analysis required to understand the subsequent components. Finally, we answer two main questions, namely (1) How does \approach decide \textit{where} to inject bugs?, and (2) \textit{how} \approach generates a set of bugs?

\subsection{Method Extractor}
\label{subsection:method_extraction}

\approach can take a single method or a repository as input. In the latter case, it first needs to 
extract the implemented methods in the project for further mutation. To that end, \textit{Method Extractor} component leverages a parser depending on the programming language used in the input project, builds program parse trees, and extracts a list of methods and constructors. Next, it collects the signature and body of the method (excluding docstrings) 
and passes them to the \textit{Attention Analyzer} component.

\vspace{-5pt}
\subsection{Attention Analyzer}
\label{subsection:where}

State-of-the-art code-language models are based on the Transformer architecture~\cite{vaswani2017attention}.
To produce contextualized vector representation of a sequence of tokens, Transformers rely on \textit{Multi-Head Self-Attention}. For a method that consists of $n$ tokens $\overrightarrow{Tkn} = \{m_0,\ldots,m_{n-1}\}$, a Transformer model with $\mathcal{L}$ layers takes $\overrightarrow{Tkn}$ as input and produces $\textbf{H}^\ell = [\textbf{h}_0^\ell,...,\textbf{h}_{n-1}^\ell]$. Here, $\textbf{H}^\ell$ corresponds to the hidden vector representations in layer $\ell \in \{1,2,...,\mathcal{L}\}$. In each Transformer layer $\ell$, multiple self-attention heads are used to aggregate the previous layer's outputs. Consequently, for each token $m_i$, the self-attention assigns a set of attention weights concerning all other tokens in the input sequence, i.e., $Attention(m_i)=\{\alpha_{i0},\dots,\alpha_{in-1}\}$, where $\alpha_{ij}$ indicates the relative attention of $m_i$ to $m_j$. 

Attention weights reflect the importance of each token in the final code representation. 
Hence, \approach 
analyze attention weights to identify tokens (and subsequently, statements) with the lowest attention weights. These statements are where \approach can change during bug-inducing transformation without impacting the overall representation of the method. Algorithm~\ref{alg:where} explains our approach for attention analysis, which takes the original method, a threshold value $k$, and a transformer-based model as inputs and pinpoints the $k\%$ of the least attended statements $LAS$ as outputs. To that end, it first extracts the list of least attended tokens in the method $LAT$ (Lines 1-10) and uses them to pinpoint the least attended statements $LAS$ (Lines 11-20).

The algorithm first identifies the tokens $\overrightarrow{Tkn}$ and statements $\overrightarrow{Smt}$ in the given method and initializes the $LAT$ and $LAS$ variables to be empty (Lines 1-3). Next, it queries the model $M$ to extract the self-attention values (Line 4). For a model $M$ with $L$ layers and $H$ attention heads per layer, the attention values will be averaged across heads and layers, resulting in an $n \times n$ matrix, where $n$ is the number of tokens. For each token $m_i$ in the $method$, the algorithm further averages the attention weight relative to other tokens (averaging the values per column in the self-attention matrix) to compute a single attention weight value for each token $m_i$ in the method (Line 5). Given that we are interested in the least attended tokens in the code, Algorithm~\ref{alg:where} sorts the attention weight vector, $\overrightarrow{TknAttnW}$, populates their corresponding indices in $\overrightarrow{SortedTknInd}$ (Line 6), and identifies the top $k\%$ of least attended tokens, $LAT$ (Lines 7-10).  

\vspace{-8pt}
\begin{algorithm}[tbh]
    \footnotesize
    \SetKwInOut{Input}{Input}
\SetKwInOut{Output}{Output}

\caption{Attention Analyzer}
\label{alg:where}

\KwInputs{
Method $method$, Threshold $k$, Transformer-based model $M$}
\KwOutput{
Least attended statements $LAS$
}

$\mathit{\overrightarrow{Tkn}} \gets \mathit{getTokens(method)}$\;
$\mathit{\overrightarrow{Smt}} \gets \mathit{getStatements(method)}$\;
$\mathit{LAT,LAS} \gets \emptyset$\;

$SelfAttnW \gets \mathit{getSelfAttnW(M, \overrightarrow{Tkn})}$\;
$\overrightarrow{TknAttnW} \gets \mathit{getTknAttnW(SelfAttnW)}$\;
$\overrightarrow{SortedTknInd} \gets \mathit{getSortedTknIndices(TknAttnW)}$\;

$\mathit{i} \gets 0$\;
\While{$i < \mathit{\lceil (k / 100) * SortedTknInd.length \rceil}$}{
    $\mathit{LAT} \gets \mathit{LAT} \cup Tkn[SortedTknInd[i]]$\;
    $\mathit{i} \gets i + 1$\;
}

$\mathit{SmtScore} \gets \emptyset$\;
\ForEach{$s_i \in Smt$}{
    $\mathit{score} \gets \mathit{| s_i \cap LAT |} / \mathit{s_i.length}$\;
    $\mathit{SmtScore} \gets \mathit{SmtScore} \cup \langle s_i,score \rangle$\;
}

$SortedSmtInd \gets \mathit{getSortedSmtIndices(SmtScore)}$\;
$\mathit{i} \gets 0$\;
\While{$i < \mathit{\lceil (k / 100) * SortedSmtInd.length \rceil}$}{
    $\mathit{LAS} \gets \mathit{LAS} \cup Smt[SortedSmtInd[i]]$\;
    $\mathit{i} \gets i + 1$\;
}
\KwRet $LAS$

\end{algorithm}
\vspace{-8pt}

With the least attended tokens extracted, the algorithm can identify the least attended statements, $LAS$. To that end, it weighs each statement by a score (Lines 11-14), which is the ratio of the number of least attended tokens in that statement, normalized by its length. The key idea here is that a statement with the highest overlap between least attended tokens should achieve a lower score and, thus, be considered the least attended statement. Without normalizing the statement length, longer statements will be penalized, i.e., \approach never selects them to mutate. Finally, the statements are sorted in ascending order based on their scores, and the least $k\%$ attended statements (we take the ceiling in case $k\%$ of total statements is less than one) will be returned as $LAS$ (Lines 15-20).

\subsection{Bug Generator}
\label{subsection:how}

Bug generation involves modifying, adding, or deleting code segments from an original method to change the expected behavior of the program. The \textit{Bug Generator} module of \approach takes a method and its corresponding set of $LAS$ identified in the previous step as inputs and crafts LLM prompts to generate $N$ buggy versions for the method as outputs. Our intuition for making \approach configurable is that our bugs will likely be used as training/fine-tuning data for bug-related tasks. So, generating multiple buggy versions of a single method would be helpful for a model to distinguish between buggy and non-buggy code more easily. The total number of bugs, however, also depends on the size of the method and threshold value $k$. For example, $k = 10\%$ for methods with less than $10$ statements returns one statement as LAS, and changing that statement in $N$ unique ways may be infeasible.

\vspace{-8pt}
\begin{algorithm}[tbh]
    \footnotesize
    \SetKwInOut{Input}{Input}
\SetKwInOut{Output}{Output}

\caption{Bug Generator}
\label{alg:how}

\KwInputs{
Method $method$, Least attended statements $LAS$, Number of bugs $N$, Transformer-based model $M$}
\KwOutput{Buggy methods $Bugs$}


$\mathit{Bugs} \gets \emptyset$\;
$\mathit{LASInd} \gets \mathit{getLASIndices(method,LAS)}$\;
$\mathit{method} \gets \mathit{addIndices(method)}$\;

$\mathit{Prompt} \gets $ "Inject \$$N$ bugs in the following method by changing only the statements at locations \$$LASInd$: \$$method$"\;



$\mathit{Responses} \gets \mathit{queryLLM(Prompt)}$\;

\ForEach{$Response \in Responses$}{
    \If{$!isParseable(Response)$}{
            $continue$\;
    }
    $newLAS \gets \mathit{getLAS(Response,M)}$\;
    $check \gets true$\;
    \ForEach{$stmt \in getDiff(Response,method)$}{
        \If{$stmt \notin newLAS$}{
            $check \gets false$\;
            $break$\;
        }
    }
    \If{$check$}{
        $Bugs \gets \mathit{Bugs} \cup \mathit{Response}$\;
    }
}

\end{algorithm}
\vspace{-8pt}

\approach's prompts consist of three parts. The first part is the natural language instruction asking LLM to generate the bugs. The second part provides contextual information about where to inject the bug, i.e., only to consider the least attended statements, $LAS$, for making bug-inducing changes. Finally, we include the entire method, including both signature and the method body, in the prompt. Such prompts are LLM-agnostic, i.e., they can be used with various existing LLMs, such as LLaMa~\cite{llama}, PaLM~\cite{chowdhery2023palm}, Copilot~\cite{copilot}, Alpaca~\cite{alpaca}, and \chatgpt~\cite{chatgpt}. The only consideration is to check if the prompt's size matches the LLM context window. After the LLM's response, \textit{Bug Generator} component validates (1) if they are syntactically correct and (2) if the changes do not impact the attention of the model. The responses that do not pass these two checks will be discarded.


Algorithm~\ref{alg:how} demonstrates \approach's bug generation and validation approach. It starts by initializing the output variable $Bugs$ (Line 1) and getting the indices of LASs in $method$ (Line 2). Next, it adds an index to each statement in $method$ (Line 3). This will help us to refer to LASs in the prompt by number instead of including the whole statements, resulting in a reduction of the prompt size. This is specifically important for longer methods or statements, as the context window of LLMs is limited~\cite{nguyen2023context,xu2022systematic}. Next, we will craft the prompt with the required context, i.e., indexed method and bug injection location, and send the prompt to an LLM (Lines 4-5)~\footnote{The natural language part of our prompts is more complex than the example here. The readers can refer to our artifact to see the exact prompts~\cite{website}}. Finally, once the LLM responds, we check if the generated bugs are syntactically correct (Lines 7-8) and if the changed statements are among the least attended statements by the model (Lines 9-14) to add them to the acceptable set of bugs (Lines 15-16).

\section{Evaluation}
\label{sec:evaluation}

To evaluate the effectiveness of \approach, we compare it with two state-of-the-art alternative approaches, namely \leam and \mubert, investigating the following research questions~\footnote{The detailed results and additional research questions, e.g., ablation study demonstrating the necessity of prompt crafting, are publicly available~\cite{website}.}:

\vspace{3pt}
\begin{enumerate}[leftmargin=*,label=\bfseries RQ\arabic*:,nosep,wide=0pt]
    
    \item \textbf{Characteristics of the Generated Bugs.} To what extent can \approach and alternative techniques successfully inject bugs into arbitrary code? What are the characteristics of the generated bugs by each approach? 
    
    \item \textbf{Effectiveness in Generating Hard-to-Detect Bugs.} How well do \textit{learning-based} bug prediction models perform at detecting \approach bugs compared to other techniques? 

    \item \textbf{Effectiveness in Generating Hard-to-Repair Bugs.} To what extent do \textit{learning-based} Automated Program Repair (APR) techniques repair \approach bugs compared to those generated by alternative approaches? 
    
    
    \item\textbf{Performance.} How long does it take to generate and validate bugs using \approach and other approaches? 
    
\end{enumerate}

\subsection{Experiment Setup and Data Availability}
\label{subsec:experimentalSetup}

\noindent \textbf{Mutant Generation.} We compare \approach with two most recent mutant generation techniques, \mubert~\cite{khanfir2023efficient} and \leam~\cite{tian2022learning}. To inject mutants, \mubert selects AST nodes representative of program behavior---literals, identifiers, expressions, assignments, object fields, method calls, array access, and types. Then, it replaces the tokens in selected AST nodes with the special token \texttt{<mask>} and uses \codebert to predict the masked token. The intuition is that if \codebert predicts a token different from the original one, the transformation introduces a bug. 
\leam is a deep learning-based technique that learns to mutate code from large examples of real-world bugs. To that end, they represent code as a sequence of AST nodes and learn to apply eight grammar rules to select and modify the code. Both \mubert and \leam claim to generate better bugs (mutants confirmed by test execution) compared to classic approaches, namely PIT~\cite{coles2016pit} and Major~\cite{just2014major}. Therefore, we did not include classic techniques in our evaluation. 

For \approach, we used the threshold value $k=10\%$, mainly because alternative approaches do not change more than a handful of statements in a given code (more details in \cref{evaluation:quality}). Our experimental results in the rest of this section confirm that even with such a low threshold, we still surpass other techniques in generating hard-to-detect (\cref{evaluation:hard-to-detect}) and repair (\cref{evaluation:hard-to-fix}) bugs. A higher threshold will make our bugs more complex compared to other approaches, improving margins. We also configured \approach to generate at most three mutants per method ($N=3$ in Algorithm~\ref{alg:how}) as our preliminary results show that with a higher number, the generated bugs for $k=10\%$ are almost identical. With a higher threshold value, one can also increase $N$ and generate more diverse bugs.
The current implementation of \approach's \textit{Bug Generator} component uses GPT-3.5-turbo due to its accessibility and effectiveness at a reasonable cost ($\$0.0004$ per mutant on average). Using a more advanced LLM such as GPT-4 likely yields better results. 

\vspace{3pt}
\noindent \textbf{Automated Bug Confirmation.} 
\approach and alternative approaches generate mutants, which entails implementing a confirmation process to ensure syntactical correctness and rule out equivalent mutants.  Following the related work~\cite{tian2022learning,khanfir2023efficient}, our confirmation procedure follows three steps: (1) running existing test suites on the original project and selecting the green tests; (2) compiling the generated mutants from the methods covered by the green tests and discarding syntactically incorrect ones; and (3) re-executing previously passed tests on syntactically correct mutants and choosing those killed through test execution. Due to the possibility of generating mutants that are not equivalent but survive test execution, we further performed a manual investigation on a subset of survived mutants. This will ensure the fairness of the experiments and the generalizability of the claims in the paper.
All the experiments were conducted with killed mutants, which we refer to as confirmed bugs. The mutant generation techniques generated a total $1,908,566$ mutants, out of which 
$699,575$
were syntactically correct, and 
$434,215$
identified as confirmed bugs through testing (details in \cref{evaluation:quality}). 

\vspace{3pt}
\noindent \textbf{Manual Bug Confirmation.} 
From the 
$265,360$
syntactically correct survived mutants, we sampled $300$, controlling equal contributions from different sets of generated mutants. We then crowd-sourced the task of \textit{mutant evaluation} to $97$ individuals with varying software engineering expertise, from beginner to expert with at least two years of industry experience. We provided each participant with the mutant and original codes and asked them to label mutants with one of \textit{Yes} (the mutant is a bug), \textit{No} (the mutant is equivalent), or \textit{Maybe} (uncertain whether the mutants is equivalent) labels. Individuals monitored their time and reported the amount spent on the task. We also asked them to evaluate how challenging it was to decide on the assigned mutants' labels. Each mutant was labeled by \textit{three} individuals, and we decided on the final label based on the majority vote. 

\begin{figure}[t]
    \centering
    \includegraphics[width=0.8\columnwidth]{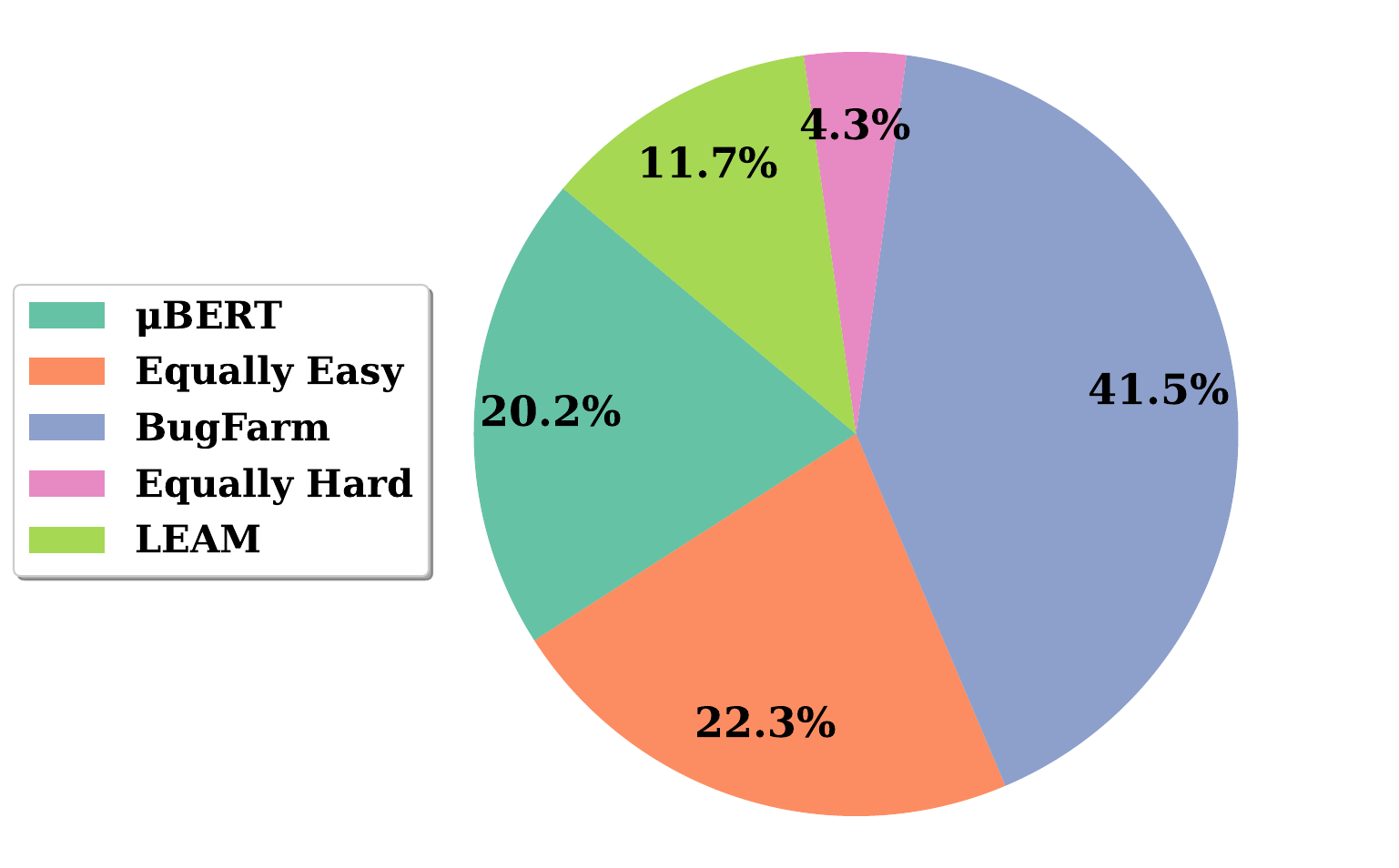}
    \caption{Evaluating the complexity of generated mutants by human subjects
    }
    \label{fig:stats}
    \vspace{-15pt}
\end{figure}

The labeling process took $290$ person-hours and participants delivered $900$ labels ($487$ Yes, $384$ No, and $29$ Maybe). The majority voting provided us with $143$ additional confirmed bugs, where $72$  $(50.35\%)$, $53$ $(37.06\%)$, and $18$ $(12.59\%)$ of them belong to \approach, \mubert, and \leam, respectively. 
During the labeling, we asked participants to identify the difficulty level for labeling each mutant, i.e., whether it was challenging for them to identify the mutant was equivalent or not. Figure~\ref{fig:stats} shows that \approach's mutants were more challenging than other approaches for the participant to label. We believe that this large-scale human study accounts for the threat to the validity of choosing killed mutants as confirmed bugs. 


\noindent \textbf{Bug Prediction Models.} When selecting bug prediction models, we had to consider the following criteria: the current implementation of \approach's \textit{Attention Analyzer} works on transformer-based models and requires the availability of models (weights and internals) to perform the attention analysis (we will consider including closed-source models as future work).
We could not find any custom bug prediction model publicly or per request available with such characteristics. Consequently, we chose three pre-trained code-language models that are widely used by the research community for fine-tuning, namely \codebert~\cite{feng2020codebert}, \codet~\cite{wang2021codet5}, and \natgen~\cite{chakraborty2022natgen}\footnote{Fine-tuning larger models requires non-trivial computing resources. Our experiments will show that models superior in other code-related tasks (e.g., \codet over \codebert) show the significance of \approach better. We expect this to hold for larger models as well.}. We fine-tuned these models using real-world bugs and evaluated the effectiveness of the outcome in predicting bugs generated by different techniques. 

\sloppy \codebert is an encoder-only transformer model based on BERT~\cite{devlin2018bert} architecture, which is trained on $2.1M$ bi-modal (natural language and code pairs) and $6.4M$ uni-modal (code only) data from CodeSearchNet~\cite{husain2019codesearchnet} dataset. The main learning objectives in \codebert are Masked Language Modeling (MLM)---the model learns to predict the tokens replaced by a special mask token---and Replaced Token Detection (RTD)---the model learns to detect which token does not belong to the original data. 
\codet is an encoder-decoder transformer model based on T5~\cite{raffel2020exploring} architecture. \codet is trained on $8.35M$ functions from various programming languages provided by CodeSearchNet~\cite{husain2019codesearchnet} and BigQuery~\cite{BigQuery} dataset, and its training objectives include masked span and masked identifier prediction. Such objectives enable the model to understand code semantics better than \codebert.

\natgen is also an encoder-decoder transformer model, trained on a generative task of naturalizing source code. Specifically, \natgen starts with \codet---based model---parameters and continues the training with a new objective, i.e., re-constructing the original code (natural) given transformed code (de-natural). It uses $8.1M$ pairs of natural and de-natural functions from CodeSearchNet~\cite{husain2019codesearchnet} and C/C\#. We used the base models of each
for our experiments.

\vspace{3pt}
\noindent \textbf{Automated Program Repair Model.} To evaluate the effectiveness of learning-based techniques in repairing generated bugs, we used \fitrepair~\cite{xia2023revisiting}, a state-of-the-art APR technique that outperforms existing approaches. It leverages information retrieval and static analysis to implement domain-specific fine-tuning and prompting strategies. Per the authors' instructions, we used \fitrepair with CodeT5-Large in a zero-shot manner to generate the patches for \mubert, \leam, and \approach confirmed bugs and validated patches through test execution.

\vspace{3pt}
\noindent \textbf{Subjects.} \approach is programming-language agnostic; none of its components depend on a specific programming language. However, we chose Java projects in our experiments for the following reasons: (1) \leam's pre-trained model is on Java, and they have no alternative training dataset for other programming languages; (2) we needed real-world bug datasets for RQ2, and most of the existing real-world bug datasets are in Java. For a fair comparison, we used Defects4J $V2.0$ projects (but used their latest version) as a baseline for bug generation since the alternative approaches are shown to work on them with no issues. The current version of \approach supports Maven projects only, so we excluded Mockito and Closure projects from the subjects. The first two columns of Table~\ref{table:quality} show the list of our $15$ subjects and the number of methods per subject used for mutant generation. 

\begin{table*}
    \setlength{\tabcolsep}{2pt}
    \centering
    \footnotesize
    \caption{Comparing the characteristics of bugs generated by \approach, \mubert, and \leam. \textbf{M}: \# Methods, \textbf{SCM}: \# Syntactically Correct Mutants, \textbf{CB}: \# Confirmed Bugs, \textbf{SI}: \# Statements Involved in bug generation, \textbf{LD}: Lines Deleted in bug generation, \textbf{ED}: Edit Distance, \textbf{OL}: Overlap with \leam, \textbf{OM}: Overlap with \mubert.
    }
    \vspace{-5pt}
    \begin{tabular}{lllllllll|lllll|lllll}
\hline
\multirow{2}{*}{\textbf{Subjects}} & \multirow{2}{*}{\textbf{M}} & \multicolumn{7}{c|}{\textbf{\approach}}                                                                                      & \multicolumn{5}{c|}{\textbf{\leam~\cite{tian2022learning}}}                                                  & \multicolumn{5}{c}{\textbf{\mubert~\cite{khanfir2023efficient}}}                                                    \\ \cline{3-19} 

                                   &                                      &       &  &     &     &     &       &      &    &  &     &   &   &       &  &       &     &    \\ [-7pt]

                                   &                                      & \textbf{SCM}      & \textbf{CB} & \textbf{SI}     & \textbf{LD}     & \textbf{ED}     & \textbf{OL}      & \textbf{OM}      & \textbf{SCM}     & \textbf{CB} & \textbf{SI}      & \textbf{LD}     & \textbf{ED}    & \textbf{SCM}       & \textbf{CB} & \textbf{SI}      & \textbf{LD}     & \textbf{ED}     \\ \hline

cli & 276 & 295 & 266 & 2.9 & 0.0\% & 33.2 & 
$\langle$13.9\%,0.6$\rangle$ & 
$\langle$8.6\%,0.7$\rangle$ & 2126 & 1388 & 2.6 & 13.69\% & 26.1 & 2053 & 1532 & 2.0 & 0.07\% & 19.3 \\

codec & 901 & 544 & 360 & 3.1 & 0.0\% & 31.6 & 
$\langle$3.1\%,0.7$\rangle$ & 
$\langle$5.5\%,0.9$\rangle$ & 4607 & 2130 & 2.8 & 7.84\% & 22.4 & 14738 & 9551 & 2.0 & 0.0\% & 17.9 \\

collections & 4440 & 3775 & 3175 & 2.9 & 0.0\% & 29.7 & 
$\langle$11.1\%,0.6$\rangle$ & 
$\langle$5.1\%,0.7$\rangle$ & 21379 & 11266 & 2.8 & 7.77\% & 20.9 & 34038 & 22196 & 2.0 & 0.0\% & 20.0 \\

compress & 4123 & 573 & 468 & 3.0 & 0.0\% & 29.4 & 
$\langle$4.4\%,0.7$\rangle$ & 
$\langle$2.4\%,0.8$\rangle$ & 20401 & 9373 & 2.6 & 12.55\% & 24.5 & 57815 & 23602 & 2.0 & 0.0\% & 24.2 \\

csv & 248 & 282 & 252 & 2.7 & 0.0\% & 24.9 & 
$\langle$11.9\%,0.7$\rangle$ & 
$\langle$1.5\%,0.7$\rangle$ & 1803 & 1189 & 2.8 & 14.55\% & 33.5 & 131 & 109 & 2.0 & 0.0\% & 7.0 \\

jxpath & 1672 & 1152 & 948 & 3.3 & 0.0\% & 28.0 & 
$\langle$8.0\%,0.7$\rangle$ & 
$\langle$6.1\%,0.8$\rangle$ & 11511 & 6169 & 2.9 & 15.24\% & 26.2 & 21335 & 10907 & 2.0 & 0.0\% & 25.8 \\

lang & 3810 & 4421 & 3791 & 3.0 & 0.0\% & 30.1 & 
$\langle$7.5\%,0.7$\rangle$ & 
$\langle$3.1\%,0.8$\rangle$ & 26365 & 16316 & 2.7 & 9.69\% & 21.4 & 23312 & 15732 & 2.0 & 0.02\% & 21.1 \\

math & 5796 & 6466 & 6207 & 3.4 & 0.0\% & 34.2 & 
$\langle$4.4\%,0.7$\rangle$ & 
$\langle$2.3\%,0.8$\rangle$ & 43090 & 42262 & 2.7 & 8.43\% & 17.0 & 85393 & 83512 & 2.0 & 0.01\% & 18.3 \\

gson & 985 & 631 & 545 & 3.3 & 0.0\% & 38.2 & 
$\langle$7.3\%,0.7$\rangle$ & 
$\langle$1.6\%,0.7$\rangle$ & 4896 & 2992 & 2.6 & 17.95\% & 27.9 & 4173 & 2067 & 2.0 & 0.05\% & 19.8 \\

jackson-core & 2626 & 2281 & 1732 & 4.0 & 0.0\% & 43.1 & 
$\langle$2.9\%,0.7$\rangle$ & 
$\langle$2.3\%,0.7$\rangle$ & 21472 & 9963 & 2.5 & 22.58\% & 25.8 & 19118 & 9310 & 2.1 & 0.0\% & 21.6 \\

jackson-db & 8076 & 4828 & 3965 & 3.9 & 0.0\% & 41.9 & 
$\langle$7.2\%,0.7$\rangle$ & 
$\langle$6.9\%,0.7$\rangle$ & 36155 & 18454 & 2.8 & 12.17\% & 28.7 & 51032 & 22454 & 2.0 & 0.0\% & 22.3 \\

jackson-xml & 586 & 312 & 258 & 3.9 & 0.0\% & 44.0 & 
$\langle$6.3\%,0.7$\rangle$ & 
$\langle$5.9\%,0.7$\rangle$ & 3096 & 1580 & 2.7 & 12.85\% & 31.1 & 2576 & 1226 & 2.0 & 0.08\% & 14.7 \\

jfreechart & 8602 & 4051 & 3186 & 3.1 & 0.0\% & 27.1 & 
$\langle$7.4\%,0.7$\rangle$ & 
$\langle$1.1\%,0.8$\rangle$ & 40313 & 17987 & 2.5 & 10.64\% & 19.8 & 18924 & 7045 & 2.0 & 0.07\% & 28.2 \\

joda-time & 4279 & 4188 & 3734 & 2.6 & 0.0\% & 22.1 & 
$\langle$8.6\%,0.6$\rangle$ & 
$\langle$3.6\%,0.7$\rangle$ & 26224 & 15641 & 2.9 & 6.29\% & 24.7 & 49907 & 28694 & 2.0 & 0.0\% & 21.1 \\

jsoup & 1642 & 1561 & 1356 & 2.8 & 0.0\% & 28.5 & 
$\langle$9.4\%,0.7$\rangle$ & 
$\langle$4.7\%,0.7$\rangle$ & 9534 & 5456 & 2.5 & 14.15\% & 23.6 & 6699 & 4013 & 2.2 & 0.05\% & 21.0 \\

\hline
\textbf{Total} & \textbf{48062} & \textbf{35359} & \textbf{30242} & \textbf{-} & \textbf{-} & \textbf{-} & \textbf{-} & \textbf{-} & \textbf{272972} & \textbf{162166} & \textbf{-}  & \textbf{-} & \textbf{-} & \textbf{391244} & \textbf{241950} & \textbf{-}  & \textbf{-}  & \textbf{-} \\

\hline
\textbf{Average} & \textbf{3204} & \textbf{2357} & \textbf{2016} & \textbf{3.2} & \textbf{0.0\%} & \textbf{32.4}& $\langle$\textbf{7.6\%},\textbf{0.7}$\rangle$ & $\langle$\textbf{4.0\%,0.7}$\rangle$&\textbf{18198}&\textbf{10811}&\textbf{2.7} & \textbf{12.43\%} & \textbf{24.9} & \textbf{26083} & \textbf{16130} & \textbf{2.0} & \textbf{0.02\%} & \textbf{20.2}

\\ \hline
\end{tabular}
    \label{table:quality}
    \vspace{-10pt}
\end{table*}

\vspace{3pt}
\noindent \textbf{Evaluation Metrics.}
To compare the performance of bug prediction models, we use accuracy, precision, recall, F1 score, 
and False Negative Rate (FNR) as our metrics. To evaluate the APR results, we measure the repair success rate, i.e., the number of bugs the technique successfully patches. \textit{True Positive (TP)} is when the code is buggy, and the model predicts it as buggy.
\textit{True Negative (TN)} is when the code is not buggy, and the model predicts it as non-buggy.
\textit{False Positive (FP)} is when code is not buggy, but the model predicts it as buggy.
\textit{False Negative (FN)} is when the code is buggy, but the model predicts it as non-buggy.


\subsection{RQ1: Characteristics of the Generated Bugs}
\label{evaluation:quality}

Table~\ref{table:quality} presents the metrics and their values calculated for the studied approaches (For \approach, the reported number is aggregated or averaged across all baseline models). 
Columns \textit{SCM} and \textit{CB} show the number of \textit{syntactically correct} mutants and \textit{confirmed bugs}, respectively. The mutant generation techniques overall generated a total of $1,908,566$ mutants, out of which $699,575$ were syntactically correct and $434,358$ confirmed as bugs (through test execution and human study). The percentage of syntactically correct to all generated mutants in \approach is $85\%$ compared to that of $59.4\%$ and $61.82\%$ for \leam and \mubert, \textbf{corroborating a higher quality of our mutants.} A higher percentage of confirmed bugs ($85.55\%$ in \approach compared to $59.41\%$ and $61.84\%$ for \leam and \mubert), along with the human study results presented in \cref{subsec:experimentalSetup}, show the code transformations by \approach are more likely to be bugs. 
The rest of the metrics are computed for confirmed bugs, which we simply refer to as bugs in the rest of the paper: 
\begin{itemize}[leftmargin=10pt]

    
    \item \textit{\# Statements involved}: This metric indicates the number of statements added, removed, or modified to generate the bugs.
    On average, even with the threshold values of $k=10\%$, \textbf{\approach changes more statements to generate bugs compared to \leam and \mubert}. Although \#SI is higher for \approach bugs, given that these statements are among the least attended statements, we will later see that the models will have a harder time distinguishing them from the correct code (more details in \cref{evaluation:hard-to-detect}). 
    
    \item \textit{\# Lines deleted}: Deleting an entire statement is very likely to
    change attention significantly and result in a runtime exception rather than a semantic bug. 
    Columns \textit{LD} show the percentage of bugs created by only deleting or commenting statements. \textbf{\approach does not delete (or comment) any statement through bug-inducing transformation ($0\%$ on average for all the projects), compared to \leam ($12.43\%$) and \mubert ($0.02$\%)}.
        
    \item \textit{Edit distance}: We also wanted to compare the edit distance between the original and generated bugs. We used Levenshtein~\cite{levenshtein1966binary} edit distance, which measures the minimum number of single-character edits---insertions, deletions, or substitutions---required to change one string into another. 
    The results for edit distance normalized by \#CB are available under columns \textit{ED}. \textbf{Compared to \leam and \mubert, \approach's bugs have higher ED values. This indicates that \approach's changes to the code are bigger yet less noticeable, as we will show in RQ2 and RQ3.} 
            
    \item \textit{Uniqueness}: We were interested to see how much our bugs overlap with those generated by \mubert and \leam. To that end, we measured the Exact Match (EM) and CodeBLEU~\cite{ren2020codebleu} values between each \approach bug with all the corresponding bugs generated by \mubert and \leam. For a given method A, if \approach generates three bugs ${b_1,b_2,b_3}$ and \leam generates four ${l_1,l_2,l_3,l_4}$, we construct $12$ pairs of $\langle b_i,l_j \rangle$ to compute the EM and CodeBLEU. EM is a strict all-or-nothing metric; being off by a single character results in a score of $0$. If the characters of $b_i$ exactly match the characters of $b_j$, EM = $1$ for the pair; otherwise, EM = $0$. CodeBLEU is a metric to measure weighted n-gram match between the pairs by considering not just the code tokens but also code syntax via abstract syntax trees (AST) and code semantics via data flow. 
    
    The $<$EM,CodeBLEU$>$ values are shown under columns \textit{OL} (overlap with \leam) and \textit{OM} (overlap with \mubert). 
    These numbers show that \textbf{only $7.6\%$ and $4\%$ of the total bugs generated by \approach for all the projects overlap with \leam and \mubert, respectively}, confirming the uniqueness of \approach bugs. 
    The high number for CodeBLEU ($0.7$ on average for both)
    shows that \textbf{although \approach bugs are better at challenging learning-based techniques (\cref{evaluation:hard-to-detect}-\cref{evaluation:hard-to-fix}), they are semantically similar to \leam and \mubert bugs; potentially as effective as them in their evaluated tasks~\cite{tian2022learning}}.
\end{itemize}

\begin{table*}[t]
    \centering
    \caption{Effectiveness of studied fine-tuned models in predicting synthetic bugs. Each set of rows shows the same pre-trained model that is fine-tuned on the same dataset but tested on different bug datasets. }
    \footnotesize

\begin{tabular}{l|cccccc}
\hline
&\textbf{Fine-tune-Model-Test}        & \textbf{Acc}  & \textbf{Prec} & \textbf{Rec}  & \textbf{F1} & \textbf{FNR}   \\ \hline
 \rownumber & \mubert-\codebert-\approach & \textbf{71.88 (4.49\% $\downarrow$)} & \textbf{89.94 (1.11\% $\downarrow$)} & \textbf{49.26 (12.18\% $\downarrow$)} & \textbf{63.66 (8.26\% $\downarrow$)} & \textbf{50.74 (15.55\% $\uparrow$)} \\
 \rownumber & \mubert-\codebert-\leam & 75.26 & 90.95 & 56.09 & 69.39 & 43.91 \\ \hdashline
 \rownumber & \leam-\codebert-\approach & \textbf{69.42 (10.76\% $\downarrow$)} & \textbf{94.05 (1.48\% $\downarrow$)} & \textbf{41.47 (28.93\% $\downarrow$)} & \textbf{57.56 (20.53\% $\downarrow$)} & \textbf{58.53 (40.53\% $\uparrow$)} \\
 \rownumber & \leam-\codebert-\mubert & 77.79 & 95.46 & 58.35 & 72.43 & 41.65 \\ \hdashline
 \rownumber & \mubert-\codet-\approach & \textbf{75.35 (0.91\% $\downarrow$)} & \textbf{87.45 (0.35\% $\downarrow$)} & \textbf{59.2 (2.18\% $\downarrow$)} & \textbf{70.6 (1.45\% $\downarrow$)} & \textbf{40.8 (3.34\% $\uparrow$)} \\
 \rownumber & \mubert-\codet-\leam & 76.04 & 87.76 & 60.52 & 71.64 & 39.48 \\ \hdashline
 \rownumber & \leam-\codet-\approach & \textbf{71.27 (8.90\% $\downarrow$)} & \textbf{89.29 (1.85\% $\downarrow$)} & \textbf{48.35 (22.86\% $\downarrow$)} & \textbf{62.73 (15.48\% $\downarrow$)} & \textbf{51.65 (38.40\% $\uparrow$)} \\
 \rownumber & \leam-\codet-\mubert & 78.23 & 90.97 & 62.68 & 74.22 & 37.32 \\ \hdashline
 \rownumber & \mubert-\natgen-\approach & \textbf{74.42 (1.29\% $\downarrow$)} & \textbf{85.07 (0.29\% $\downarrow$)} & \textbf{59.24 (3.41\% $\downarrow$)} & \textbf{69.84 (2.13\% $\downarrow$)} & \textbf{40.76 (5.40\% $\uparrow$)} \\
 \rownumber & \mubert-\natgen-\leam & 75.39 & 85.32 & 61.33 & 71.36 & 38.67 \\ \hdashline
 \rownumber & \leam-\natgen-\approach & \textbf{69.46 (9.33\% $\downarrow$)} & \textbf{88.35 (2.14\% $\downarrow$)} & \textbf{44.84 (24.80\% $\downarrow$)} & \textbf{59.49 (17.17\% $\downarrow$)} & \textbf{55.16 (36.64\% $\uparrow$)} \\
 \rownumber & \leam-\natgen-\mubert & 76.61 & 90.28 & 59.63 & 71.82 & 40.37 \\ \hdashline
 \rownumber & REAL-\codebert-\approach & \textbf{53.02 (3.84\% $\downarrow$)} & \textbf{55.23 (5.20\% $\downarrow$)} & \textbf{31.92 (11.92\% $\downarrow$)} & \textbf{40.46 (9.44\% $\downarrow$)} & \textbf{68.08 (6.78\% $\uparrow$)} \\
 \rownumber & REAL-\codebert-\mubert & 55.14 & 58.26 & 36.24 & 44.68 & 63.76 \\ 
 \rownumber & REAL-\codebert-\leam & 57.62 & 61.38 & 41.12 & 49.25 & 58.88 \\
 \hdashline
 \rownumber & REAL-\codet-\approach & \textbf{49.02 (4.72\% $\downarrow$)} & \textbf{48.93 (5.03\% $\downarrow$)} & \textbf{44.64 (9.18\% $\downarrow$)} & \textbf{46.69 (7.20\% $\downarrow$)} & \textbf{55.36 (8.87\% $\uparrow$)} \\
 \rownumber & REAL-\codet-\mubert & 51.45 & 51.52 & 49.15 & 50.31 & 50.85 \\ 
 \rownumber & REAL-\codet-\leam & 53.81 & 53.79 & 54.05 & 53.92 & 45.95 \\ 
 \hdashline
 \rownumber & REAL-\natgen-\approach & \textbf{50.75 (0.12\% $\downarrow$)} & \textbf{50.62 (0.10\% $\downarrow$)} & \textbf{61.07 (0.42\% $\downarrow$)} & \textbf{55.36 (0.23\% $\downarrow$)} & \textbf{38.93 (0.67\% $\uparrow$)} \\
 \rownumber & REAL-\natgen-\mubert & 50.81 & 50.67 & 61.33 & 55.49 & 38.67 \\ 
 \rownumber & REAL-\natgen-\leam & 53.81 & 53.01 & 67.05 & 59.21 & 32.9 \\ 
 \hline
\end{tabular}

    \vspace{-10pt}
    \label{table:cross-results}
\end{table*}

\mybox{\textnormal{\textbf{Summary.} Compared to other techniques, \approach's bug-inducing transformations involve more statement modification and change of several code tokens. The overlap between \approach bugs and other approaches is low, demonstrating their uniqueness.}}

\subsection{RQ2: Effectiveness in Generating Hard-to-Detect Bugs}
\label{evaluation:hard-to-detect}

In this research question, we investigate the effectiveness of bug prediction models on \approach bugs compared to alternative approaches. Given the unavailability of off-the-shelf models as discussed in \cref{subsec:experimentalSetup}, we fine-tuned three pre-trained code-language models (\codebert, \codet, and \natgen) using real-world and synthetic bugs. We adapted best practices for fine-tuning transformer models and used the same class distribution in fine-tuning, validating, and testing bug prediction models. All models were fine-tuned for at most $10$ epochs with loss-based early-stopping criteria of two consecutive epochs\footnote{this is the default setting for fine-tuning \codet. To ensure a fair comparison, we checked that all the models converged before $10$ epochs.}. We selected the model with the least validation error, repeated the evaluation $10$ times, and reported the average values.

Generally, one should run \approach on each model for bug generation. However, our investigation showed that fine-tuning does not greatly impact the set of LAS (Algorithm~\ref{alg:where} in \cref{subsection:where}). This is consistent with the findings of prior work that shows fine-tuning only changes the attention of the last few layers, not impacting the overall attention of the model~\cite{shi2023towards}. It also shows that many of the generated \approach bugs in this study are reusable by other researchers. Consequently, we only generated bugs for the methods per each model whose LAS set was changed compared to the baseline pre-trained model.

\subsubsection{Fine-tuning on synthetic bugs} It is possible that a learning-based bug prediction technique uses synthetic bugs for training or fine-tuning. So, we first investigate how challenging \approach's bugs are, compared to other synthetic bugs, to challenge such models. To that end, we fine-tuned the three pre-trained models on \leam and \mubert bugs. This provided us with \textit{six} fine-tuned bug prediction models, namely, \mubert-\codebert, \mubert-\codet, \mubert-\natgen, \leam-\codebert, \leam-\codet, and \leam-\natgen. We then evaluated each on \approach and the other approach, whose bugs were not used for fine-tuning. For example, we evaluated \mubert-\codebert on bugs generated by \approach and \leam. This pairwise comparison will show us which techniques generate bugs that are harder to detect. 


The rows 1--12 in Table~\ref{table:cross-results} show the result of this experiment. For Accuracy, Precision, Recall, and F1 score, the lower metric value indicates the approach's superiority (bugs are harder to distinguish from correct code, hence being detected). For FNR, a higher metric value indicates higher FN, showing the model struggles more to detect them. \textbf{From these results, we can clearly see that \approach bugs always achieve higher values for FNR (average margin of 23.31\% (min=3.34\%, max=40.53\%)) and lower values for the other metrics (average margin of Accuracy=5.95\%, Precision=1.2\%, Recall=15.73\%, and F1-score=10.84\%).}

Furthermore, F1 score values suggest the models fine-tuned on \leam have a harder time detecting \approach bugs than those fine-tuned on \mubert. We believe this is because \mubert bugs are more diverse due to changing many tokens of the code and combining them through beam search, compared to \leam bugs that try to mimic the bugs scrapped from the GitHub issue trackers. 

\begin{table*}
    \centering
    \caption{Effectiveness of \fitrepair in repairing synthetic bugs. \textbf{SI}: Statements involved.}
    \footnotesize
    \begin{tabular}{cccccc}
\hline
           & \textbf{\leam~\cite{tian2022learning}} & \textbf{\mubert~\cite{khanfir2023efficient}} & \textbf{\approach-\codebert} & \textbf{\approach-\codet} & \textbf{\approach-\natgen} \\ \hline
Total Bugs (SI=1,SI=2,SI$>$2) & 200 (174,25,1)          & 200 (187,13,0)            & 200 (125,40,35)                      & 200 (92,43,65)                    & 200 (104,30,66)                    \\ \hdashline
Success Rate     & 36\%            & 49\%              & 29\%                         & 29\%                       & 28\%                       \\ 
\hline
\end{tabular}

    \vspace{-10pt}
    \label{table:apr-results}
\end{table*}

Note that we have not used \approach bugs for fine-tuning models. The reason is \approach bugs are in-distribution samples, and alterations are within the decision boundaries for models: we only accept mutants that do not change the model's attention so that the code representation of buggy and correct code is similar (Algorithm 2 Lines 11--14). Fine-tuning on such examples theoretically cannot improve in-distribution performance and may even worsen out-of-distribution performance~\cite{kumar2022finetuning}. As we claimed before, \approach bugs are for evaluating and challenging learning-based bug-related tasks and should be used for that purpose only. 

\subsubsection{Fine-tuning on real-world bugs} To avoid any bias in our conclusion based on synthetic bugs, we also fine-tuned baseline models with real-world bugs from three datasets, namely Defects4J~\cite{just2014defects4j}\footnote{We could use the bugs from Mockito and Closure projects, which were excluded from our subjects since \approach only supports Maven at this time}, \bugswarm~\cite{tomassi2019bugswarm}, and RegMiner~\cite{song2022regminer}. The real-world evaluation dataset consists of $723$, $3285$, and $36412$ original and buggy methods from Defects4J, \bugswarm, and \regminer, respectively. This results in three bug prediction models, i.e., REAL-\codebert, REAL-\codet, and REAL-\natgen. We evaluated each model on \approach, \mubert, and \leam bugs. The rows 13--21 in Table~\ref{table:cross-results} show the results of this experiment, with margins indicating the difference with respect to second-best synthetic bug dataset. \textbf{These results show similar trends we observed with models fine-tuned on synthetic bugs, i.e., \approach bugs result in higher FNR (54.12\% on average) and lower Accuracy (50.93\% on average), Precision (51.59\% on average), Recall (45.88\% on average), and F1 score (47.5\% on average) values.}

That models fine-tuned on real-world bugs underperform those fine-tuned on synthetic bugs across all metrics. A possible justification for this observation is the distribution shift between real-world bugs (used for fine-tuning) and synthetic bugs (used for testing). The two root causes for these distribution shifts are (1) the real-world bugs belonging to projects \textit{different} than those from which we generated the synthetic bugs; and (2) the nature of real-world bugs is different from synthetic bugs. Especially for \approach and \mubert, the bug generation objective does not include any similarity to real-world bugs.

\mybox{\textnormal{\textbf{Summary.} Bug prediction models, regardless of whether the fine-tuning dataset is from a synthetic or real-world bug dataset, have a harder time detecting \approach bugs than other synthetic bugs.}}

\subsection{RQ3: Effectiveness in Generating Hard-to-Repair Bugs}
\label{evaluation:hard-to-fix}

To further demonstrate the complexity of our bugs, we evaluate the ability of \fitrepair in repairing \approach, \mubert, and \leam bugs. 
We configured \fitrepair to generate $100$ patches~\footnote{The tool often generates fewer patches than this max allowance value.} per bug and terminate if this takes more than five hours~\footnote{Confirmed by \fitrepair authors to ensure proper performance.}. Given the time-consuming nature of program repair, we evaluated \fitrepair on $1000$ generated bugs~\footnote{This number is 2.5 times than the bugs in \fitrepair evaluation.}, constitutes from $200$ sampled confirmed bugs from each bug dataset (\leam, \mubert, \approach-\codebert, \approach-\codet, and \approach-\natgen).  
When sampling, we controlled the selection of bugs from the same methods for all approaches: we sorted them based on the descending order of method size---measured by the number of characters---and picked the top $200$. 
Our rationale is that the potential locations for bug injection increase when methods are longer. As a result, the likelihood of observing different bugs produced by each technique is higher (similar to the illustrative example in \cref{sec:example}). The collected bugs are from $14$ out of $15$ subject projects. 

The first row in Table~\ref{table:apr-results} shows the distribution of generated bugs based on the number of statements involved in bug generation (\#SI from RQ1). Most subject bugs from \leam and \mubert differ in only one line with the original code, while \approach bugs are more diverse concerning this metric. Each bug dataset took \fitrepair two to seven hours to generate $35$ patches, on average. We validated all the generated patches for each bug, and if one of the patches results in a green test suite, we ignore other patches and consider the bug repaired by \fitrepair. The validation process of $1000$ bugs took over $100$ hours. The second row of Table~\ref{table:apr-results} shows the percentages of bugs that \fitrepair successfully patched. 

\textbf{\fitrepair successfully repaired 36\% and 49\% of the \leam and \mubert bugs, respectively. In contrast, it can only repair 29\%, 29\%, and 28\% of \approach bugs generated for each baseline model.} This is not surprising, as APR techniques are known to perform better in repairing bugs with SI=1. In fact, 87\% of correct patches for \mubert and \leam only differ with the corresponding bug in one line. This value is only 62\% for \approach, which implies a higher complexity of \approach's one-line bugs compared to alternative approaches. 

Our deeper investigation of the nature of bugs shows that \leam and \mubert bugs mostly change the conditional/branch statements. Since repair templates of \fitrepair are specifically designed to repair bugs in conditional/branch statements (logical expression in if statement), it can repair \leam and \mubert bugs better. In contrast, \textbf{\approach is more creative in introducing bug injection due to the power of LLMs in code synthesis, and it considers locations that the model attends less to for injecting the bugs. Consequently, it can challenge learning-based APR techniques crafted to repair known bug patterns.} By looking at the other patches that fixed bugs with SI$>$=2, we observed the same pattern, i.e., there was at least one statement changing the conditional statements. 

\mybox{\textnormal{\textbf{Summary.} When applied to the same method, \approach generates bugs that are harder to repair by learning-based repair techniques compared to alternative approaches. The power of \approach will become more evident when the methods are longer, letting it change multiple locations in the method to introduce the bug.}}

\subsection{RQ4: Performance}
\label{evaluation:performance}

\begin{figure}
    \centering
    \vspace{-5pt}
    \includegraphics[width=0.85\columnwidth]{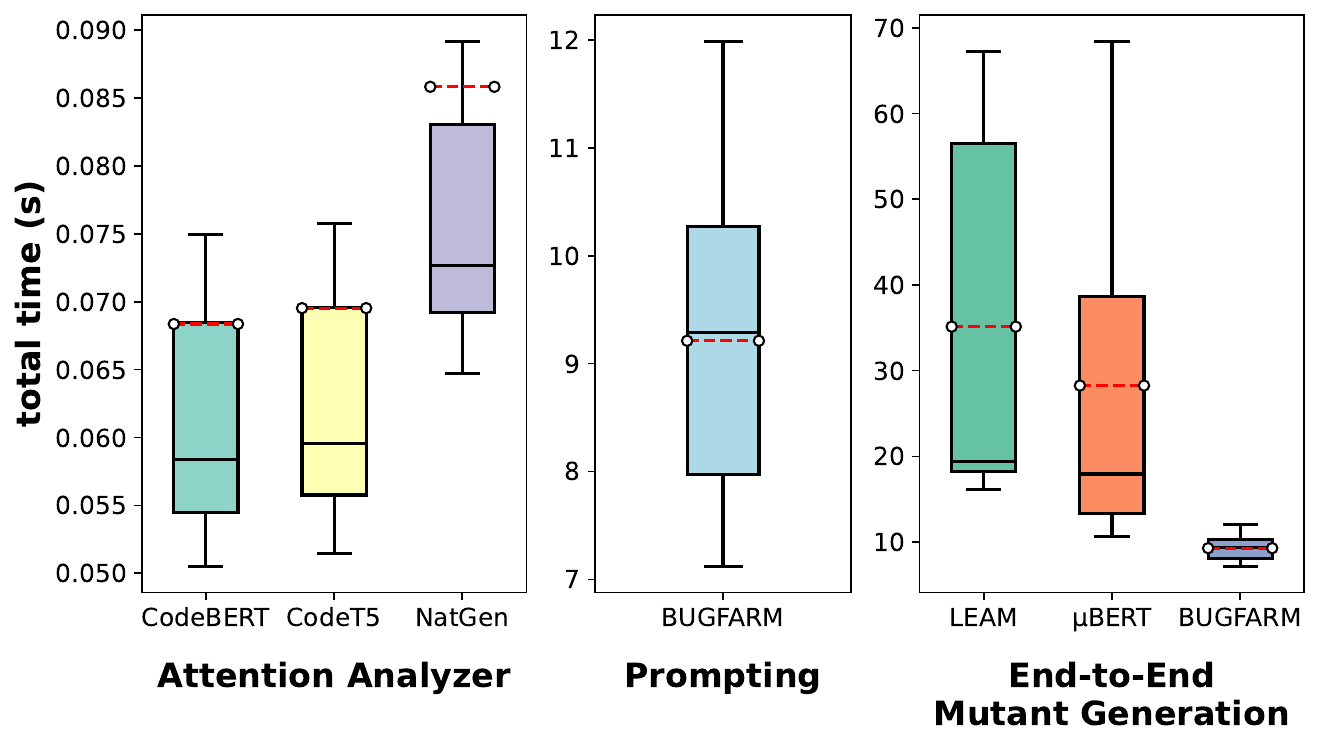}
    \vspace{-5pt}
    \caption{Performance of \approach compared to alternative approaches in mutant generation}
    \vspace{-15pt}
    \label{fig:performance}
\end{figure}

We measured the time required to extract attention weights from the models and the time it takes to prompt the LLM (GPT-3.5-turbo). We also compared the total time for bug generation in \approach compared to alternative approaches. All the experiments were performed on a workstation with NVIDIA GeForce RTX $3090$ GPUs ($24$GB GDDR6X memory) and $24$ 3.50GHz Intel 10920X CPUs ($128$ GB of memory). 
Figure~\ref{fig:performance} shows the results, where the red dashed line indicates the mean value. 
It takes $68$, $69$, and $86$ milliseconds on average for \approach to extract and analyze attention weights from \codebert, \codet, and \natgen models. Prompting the LLM for every method also takes about $9$ seconds on average (\textit{Prompting} box chart at the middle). The prompting time can be affected by multiple factors, including the traffic on the model and prompt size, i.e., the number of tokens in the prompt. 

We also measured the total time each technique takes to generate mutants (unconfirmed bugs) per method (\textit{End-to-End Mutant Generation} in Figure~\ref{fig:performance}). Compared to attention analysis and prompting, the overhead of other sub-components in \approach, such as parsing and bug selection, is negligible. \textbf{Therefore, the average time for generating all the \approach mutants per each method is $9.29$ seconds (mostly dominated by the prompting time). In comparison, it takes \leam and \mubert $35$ and $28$ seconds on average to generate mutants per method}. This is because these approaches generate more mutants, as illustrated in Table~\ref{table:quality} under \textit{\#SCM} columns. For \leam, this time does not include the training time of the model, which is $24$ hours~\footnote{This number is quoted from their paper.}. Also, since \mubert takes a long time to mutate big classes, we put a timeout of $15$ minutes on it to avoid a long generation time. 
Automated validation was also very time-consuming; we spent around $60,000$ CPU core-hours to validate over $1.9$ million mutants from \mubert, \leam, and \approach. 

\mybox{\textnormal{\textbf{Summary.} \approach is an efficient and scalable technique for generating bugs, 74\% and 67\% faster in end-to-end mutants generation than \leam and \mubert.}}

\vspace{-8pt}
\subsection{Discussion}

\approach aims to generate hard-to-detect and repair bugs concerning learning-based techniques. Our bugs are not designed to replace real-world bugs but complement them from a new perspective, i.e., having a very close code representation as the original code while different in several places. We have no claim that \approach bugs mimic real-world bugs (because they do not need to). As a result, comparing with real-world bugs is out of the scope of this paper. However, \approach leverages GPT-3.5-turbo for generating bugs, which theoretically have seen many real-world bugs during training. This can potentially help \approach bugs be similar to real-world bugs compared to \mubert, which does not concern such similarity. 
\section{Related Work}

\noindent \textbf{Real-world bug/vulnerability benchmarks.} Several attempts have been made to construct real-world bug datasets manually. Defects4J~\cite{just2014defects4j}, BugSwarm~\cite{tomassi2019bugswarm}, Bugs.jar~\cite{saha2018bugs}, and RegMiner~\cite{song2022regminer} are the commonly used Java bug datasets that have mined GitHub to collect regression bugs from bug-fixing commits. BigVul~\cite{fan2020ac} and CVEFixes~\cite{bhandari2021cvefixes} are real-world examples of vulnerabilities collected from bug-fixing reports in the CVE/NVD database~\cite{booth2013national}. \approach complements the bugs and vulnerabilities in these datasets with hard-to-repair and hard-to-detect bugs. Also, the bugs in these datasets only represent human mistakes, which could be potentially different from the mistakes AI programming tools make. 


\noindent \textbf{Learning-based bug/vulnerability generation.} Learning-based bug generation was proposed to overcome the limitations of manual defect model construction~\cite{brown2017care}. Such techniques learn the bug or vulnerability patterns from real-world bug fixes and generate mutants accordingly. DeepMutation~\cite{tufano2019learning} is a technique that relies on sequence-to-sequence neural machine translation for learning and generating bugs. SemSeed~\cite{patra2021semantic} extracts bug patterns from real-world bug fixes and injects them into other programs so that the bug in the new program is syntactically different but semantically similar. MutationMonkey~\cite{beller2021would} mines bug patterns from historical changes and transforms them into mutation operators semi-automatically. 
\vulgen~\cite{nong2023vulgen} combines pattern mining and deep learning to generate realistic bugs. 
\leam~\cite{tian2022learning} learns to mutate code from large examples of real-world bug-fixing commits. 
\mubert~\cite{khanfir2023efficient} produces buggy versions by replacing code tokens with the spacial \texttt{<mask>} token, and uses \codebert to predict the masked token. Both \leam and \mubert incorporate beam search~\cite{freitag2017beam} to generate bugs that involve more than one statement. 

\approach is superior to prior learning-based bug-generation techniques in several ways: \approach does not involve any training or fine-tuning effort to learn bug patterns and generate bugs. Consequently, it is independent of existing real-world bug datasets or a corpus of bug-fixing commits. Second, while the majority of prior work only generates one-line bugs, \approach can be configured to generate bugs that involve multiple statements. Third, \approach is the first technique that targets the generation of bugs that can challenge learning-based bug detectors and repair tools, or bugs that represent the AI programming tools' mistakes, rather than human mistakes. Our empirical evaluation confirmed that these properties result in the generation of bugs that are hard-to-detect and hard-to-repair.
\section{Concluding Remarks}

Bug benchmarks are essential in software engineering to evaluate automated techniques concerning bugs. The advent of learning-based bug-related techniques demands automated bug-generation techniques for proper evaluation. In this paper, we presented \approach, a model-in-the-loop technique for the automated generation of hard-to-detect and repair bugs. Our empirical evaluation shows the superiority of \approach to alternative mutant generation approaches in generating unique and high-quality hard-to-detect and repair bugs. \approach does not rely on existing bug datasets and is model-agnostic. 


\section*{Acknowledgments}
\label{acks}
The NSF CCF-2238045 grant supports this work. We thank Prof. Darko Marinov for sharing their wisdom to improve this research, the UIUC+'24 program students who helped with the human study and manual verification of bugs, and the Illinois National Center for Supercomputing Applications (NCSA) for providing computing resources.

\bibliographystyle{abbrv}
\bibliography{main}

\end{document}